\documentclass[12pt,twoside]{article}
\usepackage{amssymb}
\usepackage{amsmath}
\usepackage{latexsym}

\begin{document}

\title{Long-time Behavior for the Stochastic Ising Model with
Unbounded Random Couplings}

\author{H. Spohn $^{1}$, E. Zhizhina $^{2}$ }

\date{}

\maketitle

{\small \noindent

$^{1}$ Zentrum Mathematik, Technische Universit\"{a}t
M\"{u}nchen, D-80290 M\"{u}nchen, Germany, e-mail:
spohn@ma.tum.de

$^{2}$ Institute for Information Transmission Problems, Bolshoy
Karetny per. 19, 101447 Moscow, Russia, e-mail: ejj@iitp.ru
\\ }

\begin{abstract}
We consider the ferromagnetic Ising model with Glauber spin flip
dynamics in one dimension. The external magnetic field vanishes
and the couplings are i.i.d. random variables. If their
distribution has compact support, the disorder averaged spin
auto-correlation function has an exponential decay in time. We
prove that, if the couplings are unbounded, the decay switches to
either a power law or a stretched exponential, in general.
\end{abstract}

\section{Introduction and main results}

In one dimension the Ising model with spin flip dynamics has
exponentially fast mixing in time, as is reflected by the fact
that the self-adjoint generator of the stochastic dynamics has a
spectral gap, see \cite{MT} for example. One might wonder what
happens to the exponential decay when the couplings are
disordered. If the couplings are uniformly bounded, it is proved
in \cite{AMSZ,ZR} that the generator still has a spectral gap.
Thus the case of interest is when the couplings are unbounded. It
is easy to see that then the spectral gap vanishes with
probability one. The goal of our paper is to estimate how the
missing spectral gap is reflected in the decay of the disorder
averaged spin-spin correlation. In particular we will have to
identify those realizations of the couplings which are responsible
for a slow decay.

The model under the study is the one-dimensional Ising model with
formal Hamiltonian
\begin{equation}\label{H}
 H(\sigma, \omega)= - \sum_{x \in \mathbb{Z} }
\omega_{x} \sigma_{x-1} \sigma_x.
\end{equation}
Here $\sigma_{x}=\pm 1$ are the Ising spins, $\sigma\in
\Omega=\{1,-1\}^{\mathbb{Z}}$, and  $\omega_{x}$ are the
couplings. We assume that  $\{\omega_{x}, x\in \mathbb{Z} \}$ are
i.i.d. random variables with a common probability distribution
$P$. The model is assumed to be ferromagnetic, $\omega_{x}\geq 0$,
i.e. $P$ is supported in $\mathbb{R}_{+}$. The family of random
variables $\omega= \{ \omega_x, x\in \mathbb{Z} \}$ is an ergodic
random field on $\mathbb{Z}$ with the space of realizations
$\mathbb{R}_{+}^{\mathbb{Z}}$ and the probability distribution
${\mathbf{P}}=P^{\mathbb{Z}}$. It is known that for every bounded
realization of the random field $\omega$ and that for
${\mathbf{P}}$-a.e. unbounded $\omega$ the random spin system with
Hamiltonian (\ref{H}) has a unique limit Gibbs measure
$\nu_{\omega}$ for arbitrary inverse temperature $\beta$
\cite{AMSZ, ZR, GM}. To simplify our notation we include $\beta$
into the definition of coupling $\omega_x$.

For a fixed realization of couplings the Ising spin
configuration $\sigma$ evolves in time through spin flips as
specified by the flip rates
$$
c(x,\sigma, \omega)=\frac{1}{1+e^{-\Delta_x(\sigma,\omega)}},
$$
$$
\Delta_x(\sigma,\omega)= H( \sigma^{(x)},\omega) - H(
\sigma,\omega), \quad \sigma^{(x)} \in \Omega,\quad
\sigma^{(x)}_y=\left\{
\begin{array}{r}
\sigma_{y}, y\neq x, \\ -\sigma_{y}, y=x.\\
\end{array}
\right.
$$
Thus in a short time interval $ dt  $ the spin configuration $
\sigma $ changes to the spin configuration $ \sigma^{(x)} $ with
probability  $ c(x,\sigma, \omega) dt  $ and remains unchanged
with probability  $ 1 - \sum_{x} c(x,\sigma, \omega) dt  $. It is
proved in \cite{L} that this rule defines a Markov process,
denoted here by
$$
\sigma^{\omega}(t) = \{ \sigma_x^{\omega}(t), x\in \mathbb{Z}, t\ge 0
\},
$$
with state space $\Omega$. We assume that the $t=0$ distribution of
$\sigma^{\omega}(t)$ is the Gibbs measure $\nu_{\omega}$. Then $
\sigma^{\omega}(t) $ is stationary in time and reversible. The
corresponding stochastic semigroup
$T_{t}$ is self-adjoint on the Hilbert space $
{\mathcal{H}}_{\omega} = {\mathcal L}_2( \Omega, d \nu_{\omega})$.
$T_{t}$ is generated by
\begin{equation}\label{L}
 (L(\omega)f)(\sigma) = \sum_{x\in \mathbb{Z}} c(x,\sigma, \omega)
\Bigl( f(\sigma^{(x)})-f(\sigma) \Bigr),\quad f(\sigma) \in
{\mathcal D}\subset {\mathcal H}_{\omega},
\end{equation}
as acting on cylindrical functions ${\mathcal D}\subset {\mathcal
H}_{\omega}$. The operator $L(\omega)$ can be extended in
${\mathcal H}_{\omega}$ to a self-adjoint (unbounded) operator
 for $\mathbf{P}$-a.e. $\omega$ \cite{L,ZR} and will be
denoted by the same symbol. Let
$$
\lambda_0= \sup \{(L\psi, \psi),  \|\psi \| =1, (\psi, 1)=0\}
$$
denote the upper edge of the spectrum of the operator $L(\omega)$
in the subspace orthogonal to $\{1\}$. $\lambda_0$ is constant
almost surely.

The goal of our paper is to obtain the long-time behavior for the
disorder averaged time-autocorrelation function
\begin{equation}\label{00}
S(t) =   \left\langle\left\langle \sigma_0^{\omega}(t),
\sigma_0 (0)
\right\rangle_{{\mathcal{P}}(\omega)}\right\rangle,\quad
t \to \infty.
\end{equation}
Here $\langle\cdot\rangle_{{\mathcal {P}}(\omega)}$ is the average over
the the process $\sigma^{\omega}(t)$ under a fixed realization
$\omega$, and $\langle\cdot\rangle$ is the average over the
distribution $\mathbf{P}$ of random couplings. We are
interested in the case of unbounded couplings when
$$
P ( \omega_x > K ) > 0  \mbox{  for any  } K>0.
$$
In this case with probability one the operator $L$ has no spectral
gap, see for example \cite{ZR}, which implies $\lambda_0=0$.

As our main result we state  \medskip\\ {\bf Theorem 1}. {\it Let $ P
(\omega_x
>K)>0$ for every $K>0$, and
\begin{equation}\label{0}
 1< \langle (\cosh \omega_x)^{4} \rangle < \infty.
\end{equation}
Then for large enough $t$ the following estimate holds
\begin{equation}\label{T1}
 C_2 \left(\frac{ t e^{-G_{2}(t)}}{\sqrt{- g_{2}''(\mu_{2}(t))}} \right)^2
\leq S(t) \leq C_1  \left(\frac{ t e^{-G_{1}(t)
}}{\sqrt{-g_{1}''(\mu_{1}(t))}} \right)^{\frac12}
\end{equation}
with positive constants $C_1, C_2$ independent on $t$. Here
\begin{equation}
g_1(\mu) =  \ln P \left(\omega_x>\frac{1}{4} \ln
\frac{1}{\mu}\right), \quad  g_2(\mu)
=  2 \ln P \left(\omega_x>\frac{1}{2} \ln
\frac{1}{c\:\mu}\right),  \nonumber
\end{equation}
with $\mu\in (0,1)$ for suitable constant $c$,
$0<c<1$. $G_{j}$ is the Legendre transform of $g_{j}$,
$$
G_{j}(t)  =  \min_{\mu\in (0,1)} (t\mu-g_{j}(\mu)), \quad t>0,
\quad j=1,2,
$$
where the minimum is taken at $\mu_{j}(t)$.}
\medskip\\
{\bf Examples}. (i) If $P(\omega_x>u)\sim e^{-ku}$ for
$u\to\infty$ with $ k>4$, then one has
\begin{eqnarray}
 &&g_1(\mu,t)  =  \frac{k}{4} \ln \mu,\quad \mu\in (0,1),
 \quad \mu_{1}(t)=\frac{k}{4t},\nonumber \\ &&g_2(\mu,t)  =   k \ln
c\mu,\quad \mu\in (0,1), \quad \mu_{2}(t)=\frac{k}{t}, \nonumber
\end{eqnarray}
and
$$
C_2 (1+t)^{-2k} \le S(t) \le C_1 (1+t)^{-\frac{k}{8}}
$$
with constants $ C_1,C_2$ independent on $t$.\medskip\\ (ii) If
$P(\omega_{x}>u) \sim e^{-u^{\alpha}}$ for $u\to\infty$ with
$\alpha>1$, then one has
\begin{eqnarray}
&& g_1(\mu,t)  =  - \left(\frac14 \right)^{\alpha} \left(\ln
\frac{1}{\mu}\right)^{\alpha}, \quad \mu\in (0,1), \nonumber \\
&&\mu_{1}(t)  =  \alpha \left(\frac14 \right)^{\alpha} \frac{(\ln
t)^{\alpha-1}}{t} (1+ o(1)),\quad t\to \infty , \nonumber
\\ &&g_2(\mu,t)  = -  2 \left(\frac12 \right)^{\alpha} \left(\ln
\frac{1}{c\mu}\right)^{\alpha}, \quad \mu\in (0,1), \nonumber \\
&&\mu_{2}(t)  =  2\alpha \left(\frac12\right)^{\alpha} \frac{(\ln
t)^{\alpha-1}}{t} (1+o(1)),\quad  t\to \infty ,\nonumber
\end{eqnarray}
and
$$
C_2 e^{-4 (\frac12)^{\alpha}(\ln t)^{\alpha}(1+o(1))} \le S(t) \le
C_1 e^{-\frac12 (\frac14)^{\alpha}(\ln t)^{\alpha}(1+o(1))}
$$
with some positive constants $C_1, C_2$.\medskip

Our analysis estimates the integrated density of states of the
generator $L$ in the one-spin sector near zero. Using techniques
from the oscillation theorem (see \cite{GP,PF}, for example) we
establish a relation between realizations of the random couplings
and the integrated density of states. This approach is exploited
already in \cite{Z1,Z2} in the case of bounded $\omega$. There the
sub-leading correction to the exponential decay of $S(t)$ is
determined through an analysis of the asymptotics of the density
of states. For unbounded couplings, however, a new mechanism appears
resulting in a novel behavior for the spectral characteristics of
the generator. For bounded couplings the spectrum near the upper
edge comes from low probability, atypical  random couplings, for
which there are long stretches close to the maximum. They result
in a Lifschitz tail in the integrated density of states. This type
of spectrum boundary is called fluctuation boundary. In contrast,
as follows from the arguments given below, for unbounded couplings
the main contribution to the spectrum close to zero comes from
rapid oscillations of the couplings over short intervals.
 This behavior of the spectral characteristics of the generator
determines the leading decay of the disorder averaged
auto-correlation function. In particular, this implies that the
sub-leading decay for bounded couplings is unrelated to the
leading decay for unbounded couplings.

\section{Reducing subspace, proof of Theorem 1}

The auto-correlation (\ref{00}) can be rewritten as follows
\begin{equation}\label{Co}
S(t)  =  \Bigl\langle \left( e^{t L(\omega)} \sigma_0, \sigma_0
\right)\Bigr\rangle  =
 \Bigl\langle \left( e^{t L_{1}(\omega)} \sigma_0,
\sigma_0 \right)\Bigr\rangle.
\end{equation}
We must explain the meaning of the operator $L_1=L_1(\omega)$.
Since the work of R. Glauber \cite{G} it is known that
the linear span of ``one-point configurations" $\{ \sigma_x, x\in
\mathbb{Z}\}$ forms an invariant subspace for the generator
$L(\omega)$  of (\ref{L}). Moreover, the same invariant subspace
 ${\mathcal{H}}_1(\omega)\subset {\mathcal{H}}_{\omega}$ is spanned
 by the functions
\begin{equation}\label{V}
v_x(\sigma, \omega)= \cosh \omega_x \cdot \sigma_x - \sinh
\omega_x \cdot \sigma_{x-1}, \quad x\in \mathbb{Z},
\end{equation}
 see \cite{AMSZ}. The functions (\ref{V}) form the orthonormal basis in
 ${\mathcal{H}}_1(\omega)$.  We denote by $L_{1}(\omega)$ the restriction of
 the generator  $L(\omega)$ to
the invariant subspace ${\mathcal{H}}_1(\omega)$.

The operator $L_{1}$ has the following symmetric
representation in the basis $\{v_{x}, x\in \mathbb{Z}\}$
\begin{equation}\label{1}
 L_{1} v_x = A_{x, x-1} v_{x-1} + A_{x, x} v_{x} + A_{x, x+1}
v_{x+1},
\end{equation}
 with
$$
 A_{x, x-1} = A_{x-1, x} = \frac{a_x \sqrt{(1 - a^2_{x})(1 -
a^2_{x-1})} } { (1 - a_x^2  a^2_{x-1}) },\quad a_x = \tanh
\omega_x >0,
$$
$$
 A_{x, x} = -1 - \frac{a_x^2  (1 - a^2_{x-1}) } { (1 - a_x^2 a^2_{x-1})
} + \frac{a_{x+1}^2 (1 - a^2_{x})} { (1 - a_x^2 a^2_{x+1}) }.
$$
We consider new random variables
\begin{equation}\label{C}
C_{x}=  \frac{a_x^2  (1 - a^2_{x-1}) } { (1 - a_x^2 a^2_{x-1}) },
\quad C_{x} \in (0,1).
\end{equation}
Then
$$
A_{x,x-1}=\sqrt{C_{x}(1-C_{x})},\quad A_{x,x}=-1-C_{x}+C_{x+1}.
$$
Using the representation (\ref{1}) one can introduce the
integrated density of states $N(L_{1}, d\lambda)$ for the random
operator $L_{1}$ by the truncated operators
$L_{1}^{(r)}, r\in \mathbb{N}$, defined on a
finite-dimensional space of functions $ V_{r}$ of the form
$$
V_{r}= \left\{f^{(r)}(\sigma)
=\sum_{x=-r}^{r}f_{x}v_{x} \right\} \subset
{\mathcal{H}}_{1}(\omega).
$$
Let $P_{V_{r}}$ be the projection on $V_{r}$. Then the truncated
operator $ L_{1}^{(r)} =  P_{V_{r}} L_{1} P_{V_{r}}$ is given by
the same formula as (\ref{1}) when $x=-r, \ldots, r.$ We denote by
$ 0 \geq \lambda_{1}^{(r)} \geq \lambda_{2}^{(r)}\geq \ldots \geq
\lambda_{2r+1}^{(r)} $ the eigenvalues of the truncated operator
$L_{1}^{(r)}$ in decreasing order  and by $
k(L_{1}^{(r)},\lambda)$ the number of eigenvalues of $L_{1}^{(r)}$
exceeding $\lambda\in \mathbb{R}$. Then from results in \cite{PF}
it follows that there exists a non-random positive measure
$N(L_{1}, d\lambda )$ on $\mathbb{R}$, such that with probability
one
$$
\lim_{r\to\infty}\frac{1}{2r+1} k (L_{1}^{(r)}, \lambda)= N(L_{1},
\lambda)
$$
in the sense of weak convergence of measures, where
$$
N(L_{1},\lambda) = N(L_{1}, (\lambda, +\infty)).
$$
In addition
\begin{equation}\label{NE}
 N (L_{1}, \lambda) =  \left\langle \left(E_{ L_{1}}
(\lambda, +\infty )v_{0}, v_{0} \right)\right\rangle,
\end{equation}
where $\{ E_{ L_{1}} (d \lambda)  \}  $ is the spectral resolution
of the operator $L_{1}$. The representations (\ref{1}) to (\ref{C})
imply (see Lemma 1 below) that the measure $N(L_{1}, d \lambda)$
is concentrated on ${\mathbb{R}}_{-}$, so that   $N(L_{1},
\lambda) = N(L_{1}, (\lambda, 0))$  for negative $\lambda$.
\medskip\\
{\bf Main Lemma}. {\it Let $\lambda<0$ with $|\lambda|$
sufficiently small. Then
\begin{eqnarray}
&&N(L_{1},\lambda)  \geq  C_{2} \left[ P \left(
\omega_{x}>\frac{1}{2} \ln\frac{1}{c|\lambda|} \right)
\right]^{2} =  C_2 e^{g_2(|\lambda|)}, \label{L1} \\ && N(L_{1},\lambda)
\leq  C_{1} P
\left( \omega_{x}>\frac{1}{4} \ln\frac{1}{|\lambda|} \right)
= C_1 e^{g_1(|\lambda|)}
\label{L2}
\end{eqnarray}
with positive constants $ C_{j}$, $j=1,2$ and a constant $c$,
$0<c<1$. }
\medskip\\
The proof of the main lemma will be given in Sections 3 and 4 below.
We first derive the asymptotic formula (\ref{T1}) based on the
estimates (\ref{L1}), (\ref{L2}).
\medskip\\
{\bf Proof of Theorem 1}: \medskip\\ 1){\it The upper bound.}
Since, see \cite{AMSZ,Z1},
\begin{equation}\label{D}
\sigma_{x} = \sum_{y\leq x} D_{x, y}(\omega) v_{y},
\end{equation}
with
\begin{equation}\label{D1}
\begin{array}{c}
 D_{x, y}(\omega) = (1 - \tanh^{2}\omega_{y})^{1/2}
  \tanh \omega_{y+1} \ldots
\tanh \omega_{x}, \quad y<x,
\\ \\
D_{x, x}(\omega) = (1 - \tanh^{2} \omega_{x})^{1/2},
\end{array}
\end{equation}
we have
\begin{eqnarray}
&&\hspace{-55pt}\left \langle \left(e^{t L_{1}} \sigma_0, \sigma_0
\right)\right\rangle =  \sum\limits_{x\leq 0}
\sum\limits_{y\leq 0} \left \langle D_{x, 0}D_{y, 0} \left(e^{t
L_{1}}v_{x}, v_{y} \right)\right\rangle \nonumber \\ &&\hspace{20pt} \le
\sum\limits_{x\leq 0} \sum\limits_{y\leq 0} \left \langle
D^{2}_{x, 0}D^{2}_{y, 0}\right\rangle^{1/2}\left\langle \left(e^{t
L_{1}}v_{x}, v_{y} \right)^{2}\right\rangle^{1/2} \label{R1} \\
&&\hspace{20pt}
\le  \left(\sum\limits_{x\leq 0} \left \langle D^{4}_{x, 0}
\right\rangle^{1/4}\right)^{2} \left\langle \left(e^{t
L_{1}}v_{x}, v_{y} \right)^{2}\right\rangle^{1/2}. \nonumber
\end{eqnarray}
The representation (\ref{D1}) together with the condition
(\ref{0}) on the distribution of the random variables $\omega_{x}$
imply that for any $x<0$
$$
 \left \langle D^{4}_{x, 0} \right\rangle =
 \left \langle  (1-\tanh^{2}\omega_{x} )^{2} \right\rangle
 \left \langle  \tanh^{4} \omega_{x+1} \right\rangle \cdots
 \left \langle  \tanh^{4} \omega_{0} \right\rangle
 \leq \kappa^{|x|},
$$
with some $0<\kappa<1$, so that
\begin{equation}\label{R2}
\sum_{x\leq 0} \left \langle D^{4}_{x, 0} \right\rangle^{1/4}\leq
C = C(\kappa).
\end{equation}
Furthermore, for every $x,y$
\begin{equation}\label{R3}
\left(e^{t L_{1}}v_{x}, v_{y} \right)^{2} \leq \left(e^{t
L_{1}}v_{x}, v_{x} \right) \left(e^{t L_{1}}v_{y}, v_{y} \right)
\leq \left(e^{t L_{1}}v_{x}, v_{x} \right).
\end{equation}
Finally from  (\ref{Co}), (\ref{R1}) - (\ref{R3}), (\ref{NE}), and
(\ref{L2}) we conclude that for large $t$
\begin{eqnarray}
&& \hspace{-2mm} S(t) = \left\langle \left(e^{t L_{1}}\sigma_{0},
\sigma_{0} \right)\right\rangle  \leq  C^{2} \left\langle
\left(e^{t L_{1}}v_{0}, v_{0} \right) \right\rangle^{1/2}
\nonumber  \\ && \hspace{7mm} =  C^{2}\left( \left\langle \int_{R}
e^{t\lambda} \left( E_{L_{1}}(d\lambda)v_{0}, v_{0} \right)
\right\rangle \right)^{1/2} \nonumber \\&& \hspace{7mm} =
C^{2}\left( \int_{-\infty}^{0} e^{t\lambda}N (L_{1}, d\lambda)
\right)^{1/2} \nonumber  \\ && \hspace{7mm} \leq \tilde
C_{1}\left( t \int_{0}^{\infty} e^{-t\mu +g_{1}(\mu)} d\mu
\right)^{1/2}  \leq C_{1}\left( \frac{t
e^{-G_{1}(t)}}{\sqrt{-g_{1}''(\mu_{1}(t))}} \right)^{1/2},
\nonumber
\end{eqnarray}
where $g_{1}$, $G_{1}$, and $\mu_{1}(t)$ are defined in Theorem 1.
\medskip\\
2) {\it  The lower bound.} By (\ref{V}) and (\ref{0}) we obtain in
analogy with the above reasoning that
\begin{eqnarray}
&& \hspace{-5mm} \left\langle \left(e^{t L_{1}}v_{0}, v_{0}
\right) \right\rangle = \left\langle \left(e^{t L_{1}} (\sigma_{0}
\cosh \omega_{0} - \sigma_{-1}\sinh  \omega_{0}),\hspace{1mm}
\sigma_{0}\cosh \omega_{0} - \sigma_{-1}\sinh \omega_{0} \right)
\right\rangle \nonumber \\ \nonumber \\ && \hspace{10mm}
 \leq  \hspace{1mm} \left\langle \cosh^{2}\omega_{0} \left(e^{t
L_{1}}\sigma_{0}, \sigma_{0} \right) \right\rangle \hspace{1mm} +
\hspace{1mm} \left\langle \sinh^{2}\omega_{0} \left(e^{t
L_{1}}\sigma_{-1}, \sigma_{-1} \right) \right\rangle \nonumber \\
\label{R4} \\ && \hspace{-3mm} +\hspace{1mm} 2\left\langle \cosh
\omega_{0} \sinh \omega_{0} \hspace{1mm} | \left(e^{t
L_{1}}\sigma_{0}, \sigma_{-1} \right)| \right\rangle \hspace{1mm}
\leq \hspace{1mm} 2 \left\langle \cosh^{4} \omega_{0} \right
\rangle^{1/2} \left \langle \left(e^{t L_{1}}\sigma_{0},
\sigma_{0} \right)^{2} \right\rangle^{1/2} \nonumber \\  \nonumber
\\ && \hspace{2mm}  + \hspace{1mm} 2 \left\langle \cosh^{2}
\omega_{0} \sinh^{2}\omega_{0} \right \rangle^{1/2} \left \langle
\left(e^{t L_{1}}\sigma_{0}, \sigma_{-1} \right)^{2}
\right\rangle^{1/2} \hspace{1mm} \leq \hspace{1mm} k_{1} \left
\langle \left(e^{t L_{1}}\sigma_{0}, \sigma_{0} \right)
\right\rangle^{1/2} \nonumber
\end{eqnarray}
with some constant $ k_{1}$, where we used the estimate
$$
\left(e^{t L_{1}}\sigma_{0}, \sigma_{-1} \right)^{2} \leq
\left(e^{t L_{1}}\sigma_{0}, \sigma_{0} \right)  \left(e^{t
L_{1}}\sigma_{-1}, \sigma_{-1} \right) \leq \left(e^{t
L_{1}}\sigma_{0}, \sigma_{0} \right).
$$
Now from (\ref{Co}), (\ref{R4}), (\ref{NE}), and (\ref{L1}) we
derive for large $t$ the lower bound on $S(t)$ as
\begin{eqnarray}
&& \hspace{-2mm} S(t) =  \left\langle \left(e^{t L_{1}}\sigma_{0},
\sigma_{0} \right)\right\rangle \geq  k \left\langle \left(e^{t
L_{1}}v_{0}, v_{0} \right) \right\rangle^{2} \nonumber
\\ && \hspace{7mm} =   k \left( \int_{-\infty}^{0} e^{t\lambda}N (L_{1},
d\lambda) \right)^{2} \geq  \tilde C_{2}\left( t \int_{0}^{\infty}
e^{-t\mu + g_{2}(\mu)} d\mu \right)^{2}  \nonumber \\ &&
\hspace{7mm} \geq  C_{2} \left( \frac{t e^{-G_{2}(t)}}{\sqrt{-
g_{2}''(\mu_{2}(t))}} \right)^{2}, \nonumber
\end{eqnarray}
where $g_{2}$, $G_{2}$, and $\mu_{2}(t)$ are defined in
Theorem 1.
 This completes the proof of the
theorem.
$\Box$

\section{The estimate of $N(L_{1},\lambda)$ from below}

Let us fix the configuration $\omega=\{\omega_{x}, x\in \mathbb{Z}
\}$. The truncated operator $L_{1}^{(r)}(\omega)$ defined above by
(\ref{1}) is given by a Jacobi symmetric matrix of the order
$2r+1$ with positive entries $A_{x,x-1}, x=-r+1, \ldots,r$.
Consequently, for any $r$ the operator $L_{1}^{(r)}$ has
only real eigenvalues and we can exploit the technique of the
oscillation theorem in the spectral analysis for
$L_{1}^{(r)}$.
\medskip\\
{\bf Lemma 1}. {\it For every $r$ and} $ f\in V_{r}$ one has
$$
0\leq\left( - L_{1}^{(r)}f,f \right)\leq 2 \|f\|^{2}.
$$
{\bf Proof}: The proof easily follows from the obvious
inequalities
\begin{eqnarray}
 2 \sqrt{C_{y}(1-C_{y})} f_{y}f_{y-1} & \leq & (1-C_{y})
f^{2}_{y-1}+C_{y}f^{2}_{y},\nonumber \\ 2 \sqrt{C_{y}(1-C_{y})}
f_{y}f_{y-1} & \geq & -(1-C_{y}) f^{2}_{y} -
C_{y}f^{2}_{y-1}.\nonumber \quad\Box\medskip
\end{eqnarray}

Lemma 1 implies that the operators  $ L_{1}^{(r)}$ have only
negative real eigenvalues $\lambda_{j}^{(r)}(\omega), j=1,\ldots,
2r+1$. First we evaluate the function $k(L_{1}^{(r)},\lambda)$
from below for  $\lambda<0$.
\medskip\\
{\bf Definition. } We call a bond $\{x, x+1\}$  regular, if the
random variables $C_{x}$ and $C_{x+1}$, defined by (\ref{C}),
satisfy the condition
\begin{equation}\label{3}
1+C_{x}-C_{x+1}<|\lambda|.
\end{equation}
Then the following estimate holds.
\medskip\\
{\bf Lemma 2}. {\it For given } $\lambda<0$
\begin{equation}\label{4}
k(L_{1}^{(r)},\lambda) \geq {\mathcal{R}}_{r}(\lambda),
\end{equation}
{\it where $  {\mathcal{R}}_{r}(\lambda) $ is the number of
regular pairs, arranged on the interval $[-r,r]$ without
overlapping. }
\medskip\\
{\bf Proof}:  To calculate the number of eigenvalues of
$L_{1}^{(r)}$ exceeding $\lambda<0$ we will exploit the
oscillation theorem to the operator $ - L_{1}^{(r)}$ and estimate
the number $\tilde k (- L_{1}^{(r)}, |\lambda|)$
 of eigenvalues of $ - L_{1}^{(r)}$ not exceeding
$|\lambda|$: $ k( L_{1}^{(r)}, \lambda) = \tilde k (- L_{1}^{(r)},
|\lambda|)$. Let $\{f_{x}(\lambda) \}$ is an eigenfunction of $ -
L_{1}^{(r)}$ corresponding to an eigenvalue $|\lambda|$. We define
the standard phase $\varphi_{x}(\omega)$ by
$$
\mbox{ ctg } \varphi_{x+1}(\omega)=\mbox{ ctg }
\varphi_{x+1}=\frac{f_{x+1}(\lambda)}{f_{x}(\lambda)}, \quad x=-r,
\ldots, r-1.
$$
Then
\begin{equation}\label{2} \mbox {ctg }
\varphi_{x+1}=\frac{1+C_{x}-C_{x+1}-|\lambda|}{\sqrt{C_{x+1}(1-C_{x+1})}}
- \frac{\sqrt{C_{x}(1-C_{x})}}{\sqrt{C_{x+1}(1-C_{x+1})}}\cdot
\frac{1}{\mbox { ctg }\varphi_{x}}.
\end{equation}
By the oscillation theorem $\tilde k( -L_{1}^{(r)},|\lambda| )=
m_{r}(J(\lambda))+1$, where $J(\lambda)\leq |\lambda|$ is the
maximal eigenvalue of  $ - L_{1}^{(r)}$ not exceeding $|\lambda|$,
and $m_{r}(\omega, J(\lambda))$ is the number of sign changes in
the sequence of coordinates $\{f_{x}(J(\lambda))\}, x=-r,\ldots,r$
of the corresponding eigenfunction. Thus $m_{r}( J(\lambda))$
equals the number of sites $x\in [-r,r]$ with $\mbox{ ctg }
\varphi_{x}<0 $,
$$
m_{r}( J(\lambda)) = \#  \{x\in [-r,r]: \mbox{ ctg }
\varphi_{x}<0 \}.
$$
Let us consider a regular bond $\{x,x+1\}$. If  $\mbox{ ctg
}\varphi_{x}<0$, then we already have a contribution to $m_{r}(
J(\lambda))$ from that bond. If   $\mbox{ ctg } \varphi_{x}>0$,
then (\ref{2}) and (\ref{3}) imply that $\mbox{ ctg }
\varphi_{x+1}<0$. So in any case we have a contribution to $m_{r}(
J(\lambda))$ from each regular bond. Lemma 2 is proved.
$\Box$\medskip\\ Finally by averaging the inequality (\ref{4})
over realizations $\omega$ and taking the limit $r\to\infty$ we
have for $\lambda<0$,
\begin{equation}\label{5}
N(L_{1},\lambda)=\lim_{r\to\infty}\frac{\langle k(L_{1}^{(r)},
\lambda)\rangle}{2r+1} \geq b
{\mathbf{P}}(1+C_{0}-C_{1}<|\lambda|)
\end{equation}
with some constant $b$. We estimate the probability
${\mathbf{P}}(1+C_{0}-C_{1}<|\lambda|)$ under sufficiently small
$|\lambda|$ in terms of the distribution $P$ of $\omega_{x}$.
\medskip\\
{\bf Lemma 3}. {\it For all sufficiently small $|\lambda|$
\begin{equation}\label{6}
{\mathbf{P}}(1+C_{0}-C_{1}<|\lambda|)\geq p_{0} \left[ P \left(
\omega_{x}> \frac{1}{2} \ln \frac{1}{c|\lambda|}
\right)\right]^{2}
\end{equation}
 with constants $0 < p_{0} < 1$ and $0 < c < 1$.}
\medskip\\
{\bf Proof}: Let us fix some constant $h, 0<h<1$, and we denote
by
$$
p_{0} = P (0<\tanh \omega_{x}<h), \quad  0< p_{0} < 1.
$$
Then using the representation (\ref{C}) for $C_{x}$ we have for
small enough $|\lambda|$
\begin{eqnarray}
&&\hspace{-10mm}{\mathbf{P}}(C_{0}+1-C_{1}<|\lambda|)\hspace{1mm}
\geq \hspace{1mm} {\mathbf{P}}(C_{0}<|\lambda|/2; \hspace{1mm}
1-C_{1}<|\lambda|/2) \nonumber \\ \nonumber \\ && \hspace{-10mm} =
\hspace{1mm} {\mathbf{P}}\left(
\frac{a_{0}^{2}(1-a_{-1}^{2})}{1-a_{0}^{2}a_{-1}^{2}}
<|\lambda|/2; \hspace{1mm} \frac{1-a_{1}^{2}}{1-a_{0}^{2}a_{1}^{2}} < |\lambda|/2
\right) \nonumber \\ \nonumber \\ && \hspace{-10mm} \geq
\hspace{1mm} {\mathbf{P}}(a_{-1}>1-\tilde c_{0}|\lambda|;
\hspace{1mm} 0<a_{0}<h; \hspace{1mm} a_{1}>1-\tilde
c_{0}|\lambda|)\nonumber \\ \nonumber \\ && \hspace{-10mm} =
\hspace{1mm} p_{0} \left[ P (a_{x}>1-\tilde c_{0}|\lambda|)
\right]^{2} \hspace{1mm} = \hspace{1mm} p_{0} \left[ P \left(
\omega_{x}> \frac{1}{2} \ln \frac{1}{c|\lambda|}
\right)\right]^{2}\nonumber
\end{eqnarray}
with some $\tilde c_{0}$ and $0<c<1$. $\Box$ \\

The estimate (\ref{L1}) on $N(L_{1}, \lambda )$ from below follows
from (\ref{5}) and (\ref{6}).

\section{ The estimate of $N(L_{1},\lambda)$
from above}

For given $\lambda<0$ we denote by $\gamma(\lambda)=\frac{1}{4}\ln
\frac{1}{|\lambda|}$. Then for any configuration $\omega=
\{\omega_{x}, x\in \mathbb{Z}\}$ we consider a decomposition of
$\mathbb{Z}$ into two sets,
$$
\mathbb{Z}=A_{\omega,\lambda} \cup B_{\omega,\lambda}
$$
with
\begin{equation}\label{B1}
A_{\omega,\lambda}=\{x\in \mathbb{Z}:  \omega_{x}>\gamma(\lambda)\},
\quad B_{\omega,\lambda}=\{x\in \mathbb{Z}:  \omega_{x}\leq
\gamma(\lambda)\}.
\end{equation}
For any $r\in \mathbb{N}$ we denote by
\begin{equation}\label{B2}
B^{0}_{r,\omega,\lambda}=\{x\in [-r,r]:  \max
\{\omega_{x},\omega_{x-1},\omega_{x+1}\} \leq \gamma(\lambda)\}
\end{equation}
$$
 B^{0}_{r,\omega,\lambda} \subset  B_{\omega,\lambda}\cap[-r,r],
$$
and by  $W_{r,\lambda}\subset V_{r}$ the linear span of functions
$\{v_{x},\: x\in B^{0}_{r,\omega,\lambda} \}.$ Then the operators
\begin{equation}\label{25A}
 L^{(r)}_{1}=P_{V_{r}}L_{1}P_{V_{r}},\quad
L^{(W_{r,\lambda})}_{1}=P_{W_{r,\lambda}}L_{1}P_{W_{r,\lambda}}
\end{equation}
are truncations of $L_{1}$ on subspaces $V_{r}$ and
$W_{r,\lambda}$ respectively. Since  $L_{1}$ is a
self-adjoint bounded operator, we have by the minimax principle
\begin{equation}\label{k1}
k( L_{1}^{(W_{r})}, \lambda)\hspace{1mm}\geq\hspace{1mm} k(
L_{1}^{(r)}, \lambda)\hspace{1mm} - \hspace{1mm} \#\{x\in [-r,r],
x \notin B^{0}_{r,\omega,\lambda}\},
\end{equation}
where, as above, $k(A,\lambda)$ denotes the number of
eigenvalues of the operator $A$ exceeding $\lambda$.
\medskip\\
{\bf Lemma 4}. {\it  For any sufficiently small $|\lambda|$,
$\lambda<0$, and for every $\omega$ we have
\begin{equation}\label{k2}
k( L_{1}^{(W_{r,\lambda})}(\omega), \lambda)=0,
\end{equation}
where $ L_{1}^{(W_{r,\lambda})}(\omega) $ is defined in
(\ref{25A}). }
\medskip \\
{\bf Proof}: We consider the bounded configuration
$$
\tilde \omega=\{\tilde \omega_{x} \leq \gamma(\lambda), x\in
\mathbb{Z}\},
$$
coinciding with the configuration $\omega$ on
$B^{0}_{r,\omega,\lambda}$,
$$
\tilde \omega_{x}=\omega_{x}, x\in B^{0}_{r,\omega,\lambda}.
$$
Let $L_{1}^{\gamma}(\tilde \omega)$ be an operator in
${\mathcal{H}}_{1}(\tilde \omega)$ given by (\ref{1}), (\ref{C})
and corresponding to the configuration $\tilde \omega$. Our
constructions (\ref{B1}) - (\ref{B2}) imply that the operator $
L_{1}^{(W_{r,\lambda})}(\omega)$ is the same as the truncation of
the operator $ L_{1}^{\gamma}(\tilde \omega)$ on the same subspace
$W_{r, \lambda}$. As follows from results of \cite{AMSZ} in the
case of bounded couplings, under the assumption $\tilde
\omega_{x}< \gamma(\lambda)$ the upper spectrum edge of the
operator  $L_{1}^{\gamma}(\tilde \omega)$ equals to
$$
\lambda_{0}=-1+ \tanh 2\gamma(\lambda)
$$
for a.e.-configuration $\tilde \omega$, so that
$\lambda_{0}<\frac{3}{2} \lambda$ for small enough $\lambda<0.$
This estimate is valid also for any truncation of the operator
$L_{1}^{\gamma}(\tilde \omega).$ Thus no eigenvalue of the
operator $L_{1}^{\gamma,(r)}(\tilde \omega)$ or
$L_{1}^{(W_{r,\lambda})}(\omega)$ can be greater than $\lambda$.
$\Box$ \\

By (\ref{k1}) and (\ref{k2}) we have the following estimate
\begin{eqnarray}
 k( L_{1}^{(r)}(\omega),
\lambda)& \leq & \#  \{x\in [-r,r], x \notin
B^{0}_{r,\omega,\lambda}\} \nonumber \\  & \leq & C\; \# \{x\in
[-r,r], x \in A_{\omega,\lambda}\} \label{k3}
\end{eqnarray}
with some constant $C$. Applying, as before, the ergodic theorem to
the inequality (\ref{k3}), we obtain the estimate (\ref{L2}) on
$N(L_{1},\lambda)$ from above,
$$
N(L_{1},\lambda)=\lim_{r\to\infty}\frac{\langle
k_{r}(\omega,\lambda)\rangle}{2r+1} \leq  C_{1} P \left(
\omega_{x}> \frac{1}{4} \ln \frac{1}{|\lambda|}  \right).
$$


\begin{thebibliography}{15}


\bibitem{MT}
R. A. Minlos, A. G. Trishch, The complete spectral decomposition
of a generator of Glauber dynamics for one-dimensional Ising
model, Uspechi Mathem. Nauk {\bf 49},  209-210 (1994).


\bibitem{AMSZ}
S. Albeverio, R. Minlos, E Scacciatelli, E. Zhizhina, Spectral
analysis of the disordered stochastic 1-D Ising model, Commun.
Math. Phys. {\bf 204}, 651-668 (1999).

\bibitem{ZR}
B. Zegarlinski, Strong decay to equilibrium in one-dimensional
random spin systems, J. Stat. Phys. {\bf 77}, 717-732 (1994).

\bibitem{GM}
G. Gielis, C. Maes, The uniqueness regime of Gibbs fields with
unbounded disorder, J. Stat. Phys. {\bf 81}, 829-835 (1995).

\bibitem{L}
T. Liggett, Interacting Particle Systems, Springer-Verlag, Berlin,
1985.

\bibitem{GP}
S. A. Gredeskul, L. A. Pastur, Behavior of the density of states
in the one-dimensional disordered systems near the spectrum
bounds, Teoret. and Matemat. Physika {\bf 23}, 132-139 (1975).

\bibitem{Z1}
E. Zhizhina, The Lifshitz tail and relaxation to equilibrium in
the one-dimensional disordered Ising model, J. Stat. Phys. {\bf
98}, 701-721 (2000).

\bibitem{Z2}
E. Zhizhina, Spectral analysis of an one-dimensional stochastic
Ising model with random potential: asymptotics of the time
auto-correlation function, Trans. of Moscow Math. Society {\bf
64}, (2002), to appear.

\bibitem{PF}
L. Pastur, A. Figotin, Spectra of Random and Almost-Periodic
Operators, Springer-Verlag, Berlin, 1991.


\bibitem{G}
R. Glauber, Time dependent statistics of the Ising model, J. Math.
Phys. {\bf 4}, 294-307 (1963).

\end{thebibliography}
\end{document}